\documentclass[aps,prd,preprint,groupedaddress]{revtex4}

\begin{document}


\title{A New State of Baryonium}


\author{Alakabha Datta and Patrick J. O'Donnell}
\email[]{datta@physics.utoronto.ca,odonnell@physics.utoronto.ca}
\affiliation{Department of Physics,\\ University of Toronto, Toronto,
  Canada. M5S 1A7}


\date{\today}

\begin{abstract}
  
  The recent discovery of a narrow resonance in the decay $J/\psi
  \rightarrow \gamma p \bar{p}$ is described as a zero baryon number,
  ``deuteron-like singlet ${}^1S_0 $'' state. The difference in
  binding energy of the deuteron (-2.225~MeV) and of the new state
  (-17.5~MeV) can be accounted for in a simple potential model with a
  $\lambda \cdot \lambda$ confining interaction.

\end{abstract}


\maketitle

\section{Introduction}

There has been a recent observation of a near-threshold narrow
enhancement in the $P \bar{P}$ invariant mass spectrum from the
radiative decays $J/\psi \rightarrow \gamma P \bar{P}$ by the BES
Collaboration \cite{Bai:2003sw} who also report seeing nothing similar
in the decay $J/\psi \rightarrow \pi^{0} P \bar{P}$.  The enhancement
can be fit with either an S-- or P--wave Breit Wigner resonance
function. In the case of the S--wave fit, the peak mass is below $2
m_{P}$ at $M = 1859 {}^{+3}_{-10} (stat) {}^{+5}_{-25} (sys) MeV/c^2$
with a total width $\Gamma < 30 MeV/c^2$ at the $90\%$ percent
confidence level. The structure has properties consistent with either
a $J^{PC}=0^{-+}$ or $J^{PC}=0^{++}$ quantum number assignment. The
mass and width values are not consistent with any known meson
resonance near this mass.  Recently Belle has reported also
observations of the decays $B^{+} \rightarrow K^{+} P \bar{P}$
\cite{Abe:2002ds} and $\bar{B^{0}} \rightarrow D^{0} P \bar{P}$
\cite{Abe:2002tw}, also showing enhancements in the $P \bar{P}$
invariant mass distributions near $2 {m_P}$. In addition to this
probable spin zero state, there is also the report
\cite{Antonelli:1998fv} of a narrow, S-wave triplet $P\bar P$
resonance at a mass of $1870 MeV/{c^2}$ with a width of $10 MeV/{c^2}$
and $J^{PC} = 1^{--}$.

There have been some signs of an anomalous behavior in the
proton-antiproton system at a mass of $2 {m_P}$ and since the 1960's
there have been suggestions of states of nucleon- antinucleon,
sometimes called baryonium. The name has also been invoked for states
containing two quarks and two antiquarks. An example is the MIT bag
model by Jaffe \cite{Jaffe:1977ig} which postulates the existence of
baryonium for states made up of two quarks and two anti-quarks.  For a
historical review see \cite{Richard:1999qh}. In fact, the recent
observations of an unexpectedly light narrow resonance in $D^+_s\pi^0$
with a mass of $2317$ MeV by the BaBar collaboration
\cite{Aubert:2003fg}, together with a possible second narrow resonance
in $D_s\pi^0\gamma$ with a mass $2460 MeV/c^2$ have led, among other
explanations \cite{Bardeen:2003kt, Cahn:2003cw}, to a multi-quark
anti-quark model\cite{Barnes:2003dj}.  The mass difference between the
$D^*_s(2317)$ and the well established lightest charm-strange meson,
$D_s$, is $\Delta M = 350 MeV/c^2$. This is less than the kaon mass,
thus kinematically forbidding the decay $D_s^*(2317) \rightarrow
D_{u,d}+K$.  The possible resonance at $2460 MeV/c^2 $ also has such a
mass difference when taken with the lighter $D^*$ state; while this
may be a artifact of a ``feed-up'' or ``feed-down'' mechanism
\cite{Besson:2003jp} it is quite likely that both states may exist
independently.

\section{Nucleon-Nucleon and Nucleon Anti-nucleon interactions}

For over fifty years there has been a general understanding of the
nucleon-nucleon interaction as one in which there is, in potential
model terms, a strong repulsive short distance core together with a
longer range weaker attraction. Also, there have been many indications
that in the nucleon anti-nucleon system, there should be a strong
attractive $N\bar N$ bound state near threshold \cite{Shapiro:1978wi,
  Dover:1991kn}.  This understanding evolved to attribute the long
force to be that of pion exchanges and the repulsive short-range
interaction to that of $\omega$ exchange\cite{Klempt:2002ap}.  In a
nuclear physics approach this idea of meson exchange has evolved into
an accurate phenomenological way to describe experiments.

Later potential models, such as the Bonn potential \cite{Mull:1995gz},
were based on quantum chromodynamics (QCD). However, the Bonn
potential model ended up as a ten parameter model
\cite{Carbonell:1991rw} and its connection with concepts such as
one-gluon exchange are tenuous. Many of these models were based on the
non-relativistic quark model or on the MIT bag model
\cite{DeTar:1978qf, DeTar:1978pw}. The modern view is that there is a
color interaction of a $\lambda \cdot \lambda$ type between pairs of
quarks. The nucleon--nucleon or nucleon--antinucleon effective
potential then arises from the residual color forces. However to
establish the connection between the effective potential and the color
forces in practice requires somewhat {\it ad hoc} assumptions
involving either resonating-group methods \cite{Shimizu:1989ye},
variational techniques \cite{Maltman:1984st} or quark Born
perturbative methods \cite{Barnes:1993nu}.

In nuclear physics models the potential for $N\bar N$ is more
attractive than that for $N N$; this is usually considered due to
strong omega exchange which is repulsive for $N N$ and attractive in
$N\bar N$. However, the idea of $\omega$ exchange should not be taken
literally \cite{Barnes:1993nu} since there is a mismatch of the ranges
involved; $ 1/m_\omega \approx 0.2fm $ whereas the nucleon radius is
about $ 1 fm$.  In the early String/Regge approach to $N \bar N$
\cite{Chan:1978qe,Montanet:1980te} there are narrow $N\bar N$ states
based on selection rules.  For example, there is a large baryon
anti-baryon effect near threshold.  In all of these approaches there
is a trade-off between ranges - the annihilation radius is short-range
of about $1/{2M_N} \approx 0.1$fm, while the long range potentials are
dominated by meson exchanges.

\section{A simple toy model}
\label{sec:Toronto}

Within the modern QCD approach, it is the $\lambda \cdot \lambda$
color interaction that plays an important role in trying to understand
the few nucleon problem \cite{Maltman:1984st}. Nevertheless, the
actual calculational details rely on other time-honored techniques as
mentioned above. Here we wish to propose a model that has as its basis
the 6-quark state making up the deuteron.  It is known that in the
triplet neutron-proton system there is only one bound state (the
deuteron - ${}^3S_1$) with a binding energy of $-2.225$ MeV.  There is
also a large singlet scattering state, the virtual ${}^1S_0$, often
called a virtual ${}^1S_0$ state, with an energy just above zero of
$0.0382$ MeV \cite{Ma, Hulthen}. A simple phenomenological model of
the deuteron consists of using a square well potential \cite{Schiff,
  Ma, Hulthen} with a depth sufficient to bind the isoscalar ${}^3S_1$
state but not quite deep enough to bind the ${}^1S_0$ state. Then the
equation for a bound state is
 \begin{equation}
  \alpha cot(\alpha a)=-\beta \label{bound}
 \end{equation}
where 
\begin{equation}
  \alpha = \sqrt{2 M(V-E)}\label{eq2}
\end{equation}
and $\beta = \sqrt{2 ME}$, where $V$ is the depth of the potential, E,
the binding energy and $a$ the size of the well. For the deuteron $a
\approx 2 fm$. For a binding energy $E = 2.225 MeV$ the solution of
Eq. (\ref{bound}) gives a well of depth $V = 36.5 MeV$. (Here $-E$ and
$-V$ are the bound state energy and potential depth, respectively).

Our approach uses the fact the potential between two quarks due to the
$\lambda \cdot \lambda$ color interaction gives an attraction factor
of $-{2 / 3}$. In the case of $q\bar q$, the potential becomes even
more attractive by a factor of two \cite{DeRujula:1975ge,
  Maltman:1984st}.  Whether this factor of two translates into a
similar doubling of the phenomenological potential is not obvious.  We
will solve for the attractive force to fit the binding energy of the
new $P \bar{P}$ state ($17.5 MeV$). It turns out from Eq.
(\ref{bound}) the solution is $V = 64 MeV$, surprisingly just a factor
of 1.76, very close to two, deeper! Such a stronger attractive force,
such as that expected from the color factor in the potential would
seem to be consistent with the new $0^{-+}$ being a real ${}^1S_0$
bound state. In the case of the deuteron, we might assign the role of
hyperfine interactions to raise the effective potential from that
which binds the ${}^3S_1$ ``deuteron'' to a value which just fails to
bind the virtual ${}^1S_0$ state. For $N \bar N$ baryonium we expect
that the annihilation is a short range phenomenon, which can modify
the affect of the short range hyperfine interactions making its role
in the $P \bar{P}$ system unclear.  Hence there is no simple way to
predict the potential for the ${}^3S_1$ $P \bar P$ state. Also, while
there is a clear distinction between the spin-one $PN$ deuteron
(${}^3S_1$) being isoscalar and the spin-zero ${}^1S_0$ being
isovector, no similar distinction can be made for the nucleon
anti-nucleon state since both $I=0$ and $I=1$ states can exist with
either spin-zero or spin-one \cite{Gibson}

However, this does not mean they should all be seen in the $J/\psi
\rightarrow \gamma P \bar{P}$ as we show in table 1.
\begin{table}
\caption{Final states that are allowed or disallowed: here, I, C mean that
  the final states are disallowed by isospin or charge conjugation \label{1}}
\begin{ruledtabular}
\begin{tabular}{llll}
Final State in $J/\psi \rightarrow \gamma P\bar P$ & Isospin ($P \bar P)$& $J^{PC} $ &Allowed\\
$\gamma + {}^1S_0$ & 0 & $0^{-+}$ & yes\\
$\gamma + {}^1S_0$ & 1 & $0^{-+}$ & no (I)\\
$\gamma + {}^3S_1$ & 0 & $1^{--}$ & no (C)\\
$\gamma + {}^3S_1$ & 1 & $1^{--}$ & no (C,I)\\
$\pi^0 + {}^1S_0$ & 0, 1 & $0^{-+}$ & no (C)\\
$\pi^0 + {}^3S_1$ & 0 & $1^{--}$ & no (I)\\
$\pi^0 + {}^3S_1$ & 1 & $1^{--}$ & yes (OZI suppressed)\\
\end{tabular}
\end{ruledtabular}
\end{table}

\section{Summary and Conclusions}

We have presented a simple model where we account for the new $0^{-+}$
state of $P \bar P$ being a bound state of baryonium comparable to the
${}^1S_0$ virtual bound state of the deuteron. This would imply that
the ``deuteron'' equivalent ${}^3S_1$ state may also exist although we
do not have any guidance on how to derive the size of the equivalent
potential. The $1^{--}$ state at $1870 MeV/c^2$ seen in the $
e^{+}e^{-} \rightarrow P \bar P$ would appear to be a suitable
candidate. It also would appear likely that similar types of baryonia
should exist; for example, a calculation similar to that described
above but with the $\Lambda$ mass substituted for that of the proton
predict another $\Lambda \bar \Lambda$ $0^{-+}$ state with a binding
energy of $31MeV$ at a mass of $2200 MeV/c^2$. Therefore, the idea
that these resonances could be analogous to the ``virtual bound
state'' in the N-P system implies that further resonances should be
expected.  Rosner has also looked at baryon anti--baryon enhancements
in B decays \cite{Rosner:2003bm}.  He also notes that there is a whole
new interesting set of B decays possible involving exotic mesons and
baryons.
 
\section{Acknowledgments}

We thank Steve Olsen for bringing this new effect to our attention and
for discussions. This work is supported by the Natural Science and
Engineering Council of Canada (NSERC) under grant number A3828.

\end{document}